\documentclass{aa}  
\usepackage{graphicx}
\usepackage{txfonts,natbib}
\begin{document}
   \title{Dynamical evidence of the age--metallicity relation in the Milky Way disk\thanks{This
           paper is based on data collected at the Observat\'orio do Pico dos Dias, operated by Laborat\'orio
           Nacional de Astrof\'{\i}sica, MCT, Brazil.}}

   \author{H. J. Rocha-Pinto\inst{1}
          \and R. H. O. Rangel\inst{1}
          \and G. F. Porto de Mello\inst{1}
          \and G. A. Bragan\c ca\inst{1}
         \and W. J. Maciel\inst{2}
           }

   \offprints{H. J. Rocha-Pinto}

   \institute{Observat\'orio do Valongo, Universidade Federal do Rio de Janeiro,
              Ladeira do Pedro Ant\^onio 43, 20080-090 Rio de Janeiro RJ, Brazil.
              \and 
              Instituto de Astronomia, Geof\'{\i}sica e Ci\^encias Atmosf\'ericas, Universidade de S\~ao Paulo, 
              R. do Mat\~ao 1226, 05508-900 S\~ao Paulo SP, Brazil
             }

   \titlerunning{Dynamical evidence of the age--metallicity relation}
   \authorrunning{H. J. Rocha-Pinto et al.}

   \date{Received date; accepted date}

 
  \abstract
   {}
   {We studied the relationship between the average stellar abundance of several elements 
      and the orbital evolution of stars in the neighbourhood of the Sun. }
   {We used both observational data for 325 late-type dwarfs in 
    a volume-complete sample and simulations of the orbital diffusion. Metallicities, ages, and initial 
    position and velocities for the simulated stars are sampled from empirical distributions of these 
    quantities in the Milky Way.}
   {We found that that there is a relationship between the average stellar abundance of Fe, Na, Si, Ca, Ni, and Ba 
    and the mean orbital radius of stars currently passing through the solar neighbourhood. The greater the difference 
    between the mean orbital radius and the solar Galactocentric distance, the more deficient the star is, on average, 
    in these chemical species. }
   {The stars that take a longer time to come from their birthplaces to arrive in the present solar neighbourhood are more 
    likely to be more metal-poor than those that were born here. This result is a direct, independent indication 
    that a tightly defined Galactic age-metallicity relation exists.}

      \keywords{Galaxy: evolution -- stars: late-type -- solar neighbourhood}

   \maketitle

\section{Introduction}

Although one of the core consequences of the chemical evolution
theory is the gradual increase in the metal content of the
interstellar medium (ISM) and the
progressive enrichment of subsequent stellar generations, some authours 
have found little, if any, indication that
an age--metallicity relation (AMR) exists amongst solar
neighbourhood late-type stars (\citealt{feltzing,bruxa}). 
None of these papers question the predictions of the chemical evolution
theory concerning the enrichment of the ISM. The discrepancy
between theory and data is usually interpreted as a result of an
inefficient mix of the stellar ejecta into the surrounding medium, in spite of 
the growing evidence that the present dispersion in the ISM over distances of 1 
kpc is very small 
\citep{scaloelmo}. However, other authors have found a much tighter
relationship between age and metallicity (see \citealt{r00}). 

A way out of
this deadlock was recently proposed by \citet{pont}, who showed
that the non-linear separation of the isochrones in the HR diagram
can introduce a statistical bias in the isochrone age measurement,
artificially assigning a greater isochrone age to the stars.
\citet{pont} showed that if a proper statistical method is used to
correct for this bias, a tight AMR can be found from the same data
used by some of the `loose' AMR proponents. On the other hand, it
is true that chromospheric ages also suffer from their own biases
and cannot be considered perfect age estimators
\citep{donahue,pace}. Since much of the problem stems from the
drawbacks of age determination methods, the existence of an AMR
can be tested with the use of an age-related quantity that is
independent of any age determination method. In this Letter, we
show that the mean orbital radius of stars presently passing
through the solar neighbourhood, $R_m$, is mildly related to the
stellar age and that it can provide this independent test of the
existence of the AMR.

\section{Sample selection and observations}

Our sample consists of stars from the Hipparcos catalogue
\citep{hipparcos} that also have {\it uvby} indices measured by
\citet{olsen83,olsen93,olsen94} or compiled by \citet{HM98} and
which could be observed from the southern hemisphere. A
heliocentric distance cut of 25 pc was used in order to have a
sample of nearly 300 stars, so it would be possible to complete
the programme within a few observation nights at a small telescope.
Our final sample has 325 stars.

The observations were carried out at the Laborat\'orio Nacional de
Astrof\'{\i}sica, as part of a survey of the chemical abundances of
late-type dwarfs, during 15 nights between July 1999 and October
2001. The coud\'e spectrograph of the 1.60m telescope was used
with the 1800 l/mm grating and a 1024-pixel CCD, pixel size being
24$\mu$. The spectra were centred at $\lambda 6145\AA$, covered about
150 $\AA$, and had a nominal resolution of 0.3 $\AA$, with an
average S/N ratio of 100 to 200. All spectra were reduced in the
standard way; continuum normalization was performed homogeneously
for all spectra. The spectral range contains a number of
moderately strong, unblended transitions: 2 for Na I, 3 for Si I,
3 for Ca I, 10 for Fe I, 2 for Fe II, 5 for Ni I, and one for Ba
II. The equivalent widths (EWs) of the metal lines were converted
to elemental abundances using the MARCS model atmospheres
described by \citet{edv}, in a differential analysis with the Sun
as standard star. Solar $gf$ values were derived by forcing EWs
measured off Moon spectra, obtained in the same conditions as
the stellar spectra, to reproduce the standard solar abundances.
Effective temperatures for the program stars were derived from
$(b-y)$, $(B-V)$ and $(B_{\rm T}-V_{\rm T})$ calibrations described
by \citet{delpelosoetal2005}. Surface gravities were obtained from
the ionization equilibrium of the Fe I and Fe II lines. A full
description of the abundance analysis will be presented elsewhere.
The internal uncertainty of the [X/H] abundances is $\sim$~0.1 dex,
except for Ba II, which has more uncertain abundances, with only 
one spectral line measured.

Radial velocities ($v_{\rm r}$) for most of the 325 observed stars
were calculated from Th-Ar comparison spectra. For a few of them,
comparison spectra could not be obtained, and their $v_{\rm r}$ 
were taken from the SIMBAD database. We calculated the $U$,
$V$, and $W$ components of the spatial velocity around the Galactic
centre, from the $v_{\rm r}$, Hipparcos parallaxes and proper motions. 
Thereafter, the orbital parameters (mean orbital
galactocentric radius, eccentricity, maximum height above the
plane) for each star were calculated as described in
\citet[][hereafter, R04]{r04}. In this paper, we have adopted the
present solar Galactocentric radius, $R_\odot$, as 8.0 kpc.

\begin{figure}
      \resizebox{\hsize}{!}{\includegraphics{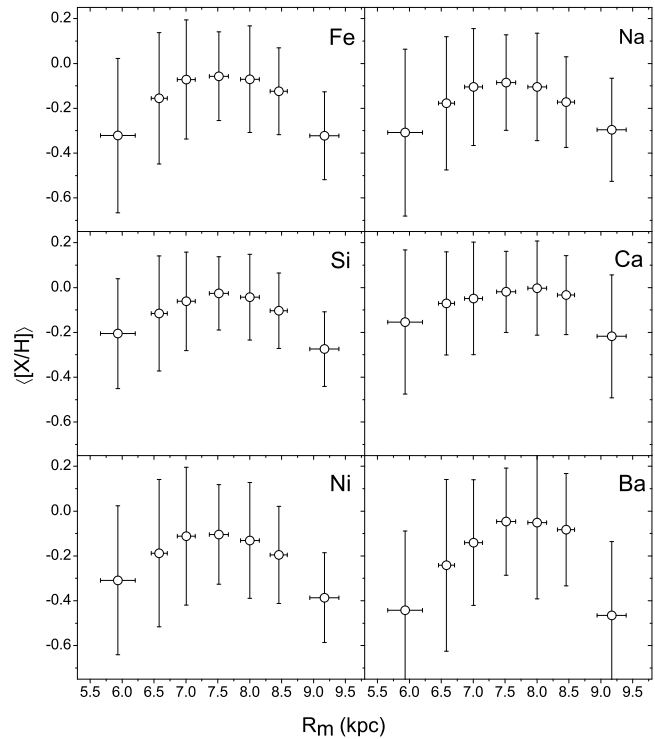}}
      \caption[]{Average abundance of Fe, Na, Si, Ca, Ni, and Ba as a function of the mean orbital Galactocentric radius for stars
            presently crossing the solar neighbourhood.
            The average abundance is larger for stars having $R_m$ around the present solar Galactocentric radius, and it decreases by
            more than 0.2 dex for stars at extreme $R_m$. The vertical and horizontal error bars are standard 
            deviations around $\langle[{\rm X/H}]\rangle$ and $\langle R_m\rangle$, respectively, at each bin.}
       \label{ztrend}
     \end{figure}

A full chemokinematical analysis of this data sample will be
presented elsewhere, but here, we are interested in an apparent trend
between $R_m$ and the stellar abundances. Figure \ref{ztrend}
shows plots of average elemental abundance, [X/H], averaged over
several $R_m$ bins, for Fe, Na, Si, Ca, Ni, and Ba. According to
\citet{grenon} and \citet{edv}, $R_m$ can be taken as an indicator
of the stellar Galactocentric birth radius, $R^\ast$. \citet{wfd}
have shown that this relation between age and $R_m$ can only be
considered in a statistical sense. However, \citeauthor*{r04}
found a correlation between the age and $R_m$ range for the stars
presently crossing the solar neighbourhood: younger stellar
generations have $R_m$ near $R_\odot$, while the older generations
have the broader $R_m$ distributions typical of stars coming from
different parts of the Galaxy into the solar neighbourhood. This
finding is consistent with Grenon's proposition that $R_m$ stays
more or less close to $R^\ast$ and reinforces previous claims that
$R_m$ can be used to estimate $R^\ast$.

In Fig. \ref{ztrend}, we show that the average abundances for
stars presently crossing the solar neighbourhood also differ as a
function of $R_m$. The stars with $R_m \approx R_\odot$ have
near-solar $\langle[{\rm X/H}]\rangle$, while stars supposedly
born farther and farther from the solar neighbourhood (i.e., those
with larger $|R_m-R_\odot|$) have progressively decreasing
$\langle[{\rm X/H}]\rangle$. This decrease can be as high as 0.3
dex for Fe and 0.4 dex for Ba, between $R_m=8$ and $R_m=6$ kpc. 

An anonymous referee has pointed out that the same trend can be seen 
in seen in Table 14 of \citet{edv}; we have also found it for [Me/H] amongst 
late-type dwarfs from the Geneva-Copenhagen survey (see 
Fig. \ref{result}).

Disk stellar velocities peak near $\Omega(R^\ast)$, the average
rotational velocity at the stellar birth radius. Most of this
motion goes in the azimuthal direction, not in the radial
direction. Radial and vertical movement is mostly driven by the
orbital diffusion \citep{wielen}, on account of stellar encounters
with massive objects like giant molecular clouds. As 
time goes on, the distance between the apogalactic and perigalactic 
radius increases due to the random kicks that 
stars get from collisions, and the chance the star could be caught
at a different galactocentric distance from its birth radius
increases as well. Thus, since $R_m$ can be seen as a stellar
birth radius indicator, we expect that for our local sample
$|R_m - R_\odot|$ reflects the time the star spent to come from
$R^\ast$ to $R_\odot$. In other words, $|R_m - R_\odot|$ can be
taken as an independent age estimator. It has an obvious advantage
of neither depending on stellar evolution models nor being based on
chromospheric properties. However, like other kinematical age
estimators, it does not allow us to assign ages for individual
stars, but to a stellar group. This is easily seen from Figure 7
by \citeauthor*{r04}: The larger the age spectrum, and consequently the age
uncertainty, the smaller $|R_m - R_\odot|$. Hence, while
stars with $R_m\approx R_\odot$ have a wide age range, stars
that came into the solar neighbourhood from other radii are, on
average, older than the stars born near $R_\odot$.

If this is the case, the relation between average abundance and
$R_m$ seen in Fig. \ref{ztrend} reflects the existence of a
relation between age and elemental abundances that is similar to the AMR;
otherwise, if there were no AMR in the Galactic disk, as hinted
by \citet{feltzing}, solar neighbourhood stars with large $|R_m -
R_\odot|$ (old stars according to the reasoning presented in 
the previous paragraph) could have any abundance within the typical 
abundance range of disk stars, and their average would not be dissimilar 
to the average abundance of stars having $|R_m - R_\odot|\approx 0$.

\section{Simulation of the $R_m$--abundance plane}

To test this hypothesis, we ran chemodynamical
simulations of a stellar sample, using a reasonable Galactic model
constrained by empirical data. Because we are interested in thin
disk stars, we did not address the simulation of the thick disk or
halo, although the effect of these Galactic components on the
Galactic potential is taken into account.

The starting position of a simulated star in $X$, $Y$, and $Z$
coordinates with respect to the Galactic centre is randomly chosen
following the thin disk density law of \citet{siegel}. The star
formation rate was considered to be constant, and a randomly uniform age
from 0 to 15 Gyr is assigned to each star. The simulated stellar
metallicity follows the distribution
\begin{equation}
{\rm [Fe/H]} \sim N\left(z(\tau) + (R-R_\odot)\partial_{R}{\rm [Fe/H]},\sigma^2\right),
\label{metsim}
\end{equation}
where $z(\tau)$ represents the AMR at age
$\tau$, $\sigma$ is the cosmic scatter, and $\partial_{R}{\rm
[Fe/H]}$ the radial metallicity gradient at the galactocentric
distance $R$. Finally, the starting velocities $V_r$, $V_\phi$, and
$V_z$ for each star, in the radial, azimuthal, and vertical
direction, respectively, with respect to the Galactic centre, were
sampled from a normal distribution $\sim N\left(\langle
V_i\rangle, \sigma^2_{V_i}\right)$, where $\langle V_r\rangle =
\langle V_z\rangle = 0$ and $\langle V_\phi \rangle = \Theta(R)$,
i.e., the rotation velocity at $R$, taken from \citet{law}. We took
$[\sigma_{V_r},\sigma_{V_\phi},\sigma_{V_z}]=[23,12,9]$ km/s from
\citeauthor*{r04}. 

The orbit of each star was then followed from its birth ---
given by the randomly assigned age --- to the present time, using
a program developed by Chris Flynn (details in \citeauthor*{r04}).
To simulate stellar encounters with massive objects, a random push
in each velocity component was given at each integration step. The
magnitude of the push in each velocity component was set by
requiring that the increase in the velocity dispersion with age of
the stellar group for the simulated stars in the solar vicinity
fitted the observed age--velocity dispersion relation (given by \citeauthor*{r04}).  
After integrating the orbit, stars whose present-time galactocentric
radius differed by less than 0.5 kpc from $R_\odot$ were selected
as ``solar neighbourhood'' stars. It is from this simulated
subsample of stars passing through the solar neighbourhood that we
take the predicted $R_m$--abundance relation to compare with the
results from Fig. \ref{ztrend}.

    \begin{figure}
      \resizebox{\hsize}{!}{\includegraphics{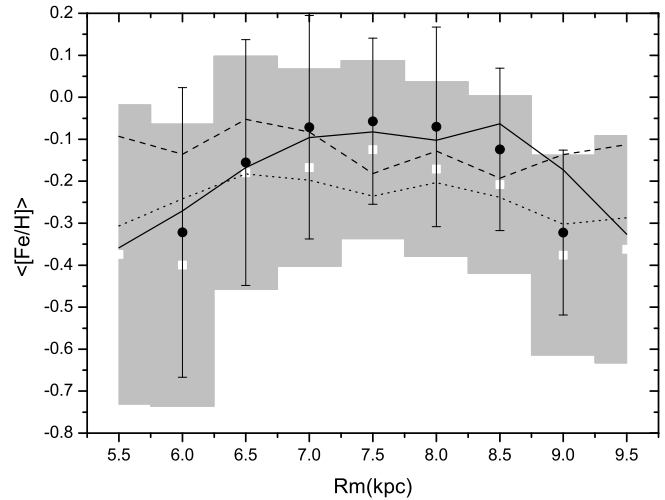}}
      \caption[]{Comparison between the observed and predicted 
                 $R_m$--[Fe/H] trend for stars passing through
                  the solar neighbourhood.  The hatched grey area and 
                  white squares correspond to standard deviations around the 
                  average [Fe/H] for the same $R_m$ intervals for 13017 
                  late-type dwarfs in the Geneva-Copenhagen survey \citep{bruxa}. 
                  The lines correspond to 
                  the results of a Galactic model with no AMR (dashed line), 
                  loose AMR (dotted line), and
                  tight AMR (solid line), as detailed by the parametrizations 
                  in Table \ref{paramet}.}
       \label{result}
     \end{figure}

Since our goal was simply to show that the observed trend in the
$R_m$--abundance plane is determined by the existence of a
well-defined AMR, we used three single parametrizations for the
AMR: i) pure scatter relation, for which no change in the
average metallicity or abundance scatter occurs with time; ii) a loose 
AMR, corresponding to a polinomial fit to the data in \citet{edv}; 
iii) a relatively tight relation according to which the average
metallicity of younger stars is higher than that for older stars.
The adopted parametrizations for these two cases are given in
Table \ref{paramet}. 

     \begin{table}
      \caption[]{Parametrizations used in the simulations}
         \label{paramet}
         \begin{flushleft}
      \begin{tabular}{lc}
         \hline\hline
            function & parametrization \\
         \hline\hline
            no AMR & $z(\tau)=-0.15$; $\sigma=0.3\,{\rm dex}$ \\
          loose AMR & $z(\tau)=0.0239 - 0.0355\tau - 0.0007\tau^2$ \\
                        & $\sigma=0.22\,{\rm dex}$ \\
      \strut  tight AMR & $z(\tau)= 1.51 - 44.25(34.23-\tau)^{-1}\strut$ \\
                        & $\sigma=0.1\,{\rm dex}$ \\
         \hline
            single-slope $\partial_{R}$[Fe/H] & $-$0.07 dex kpc$^{-1}\phantom{, \quad R < R_\odot}$ \cr
            two-slope $\partial_{R}$[Fe/H] & $-$0.05 dex kpc$^{-1}, \quad R \le R_\odot $\cr
                                           & $-$0.03 dex kpc$^{-1}, \quad R > R_\odot$ \cr
         \hline\hline
      \end{tabular}
      \end{flushleft}
   \end{table}

Our simulations consider an initial number of 6000 stars, in order
to have nearly 300 stars within 0.5 kpc from $R_\odot$ at the end
of the integration to compare with the observational sample.
Fig. \ref{result} summarises our simulations of the
$R_m$--abundance plane. 
Three curves are shown, which correspond to
the $R_m$--$\langle{\rm [Fe/H]}\rangle$ trend found using the tight AMR, loose AMR, 
 and no--AMR case, each with the two-slope metallicity
gradient given in Table \ref{paramet}. 
Clearly the no--AMR
simulation failed to produce the arc-shaped $R_m$--$\langle{\rm [Fe/H]}\rangle$ trend
seen in the observational data. This failure is not difficult to
understand, because if there is no AMR in the Galactic disk, there is no
relation between how much time a star takes to come from $R^\ast$
to $R_\odot$ and its metallicity. The $R_m$--$\langle{\rm [Fe/H]}\rangle$ trend for this
case is close to the adopted average metallicity for this
parametrization ([Fe/H] $\approx -0.15$ dex). The trend also
decreases very slowly with $R_m$ on account of the adopted
$\partial_{R}{\rm [Fe/H]}$.

The solid curve in Fig. \ref{result} is very similar to the
observed trend. It is the resulting $R_m$--$\langle{\rm [Fe/H]}\rangle$ trend for the
case of a tight AMR with very small intrinsic metallicity
dispersion. We have made no attempt to find a best-fit
curve, since the goal of the simulations is to show
that the observational trend between $R_m$ and [Fe/H] can be
explained with reasonable assumptions for the disk parameters,
provided that there is a tight AMR. Several other parametrizations
for the AMR and metallicity gradient were used, in order to verify
the effect of the parametrization on the $R_m$--$\langle{\rm [Fe/H]}\rangle$ trend. We
verified that the shape of the resulting curve is
most sensitive to the overall metallicity growth from the
beginning of the Galactic disk to the present time, as well as to
$\sigma$. A lower overall metallicity growth and/or larger
$\sigma$ flatten the simulated $R_m$--$\langle{\rm [Fe/H]}\rangle$ relation, making it look
closer to the no--AMR case. On the other hand, the effect of
$\partial_{R}{\rm [Fe/H]}$ is more pronounced on the tails of the
$R_m$--$\langle{\rm [Fe/H]}\rangle$ curve. We verified that a two-slope metallicity
gradient \citep{maciel} yields a better eye-ball fit to the observed trend than
a single-slope gradient. However, the effect is small and should
be investigated more before any conclusions can be made.

Our simulations have only considered the $R_m$--$\langle{\rm [Fe/H]}\rangle$ trend
because there is much more information available in the literature
for the observed relations between [Fe/H], $\tau$, and $R$. However,
the same exercise could be done for Ca, Na, Si, Ni, and Ba, as long
as their abundances with respect to H have grown since the
beginning of the Galactic disk and the intrinsic cosmic scatter at
any age was low. It is interesting to verify in Fig.
\ref{ztrend} that the biggest difference in the abundance at
$R_m\approx R_\odot$ and large $|R_m-R_\odot|$ is found for Ba.
Barium is thought to present a larger overall growth with age than
iron \citep{edv}, since the bulk of its nuclesynthesis is due
to long-lived AGB stars, which make their presence felt in the
chemical evolution of the Galaxy with a lag with respect to iron 
sources. From our
simulations, a larger growth in abundance with age increases the
arc-like shape of the $R_m$--$\langle{\rm [Fe/H]}\rangle$ relation, on account of the
difference in abundance from the older stars with larger
$|R_m-R_\odot|$ to the average abundance stars with $R_m\approx
R_\odot$. It is also interesting to note that Ca and Si (and
perhaps also Na) seem to have the flatter profiles with $R_m$ in
Fig. 1. That they are $\alpha$-elements, almost exclusively
products of type II supernovae, could account for this, since a less 
well-defined age--abundance relation is expected in this case.

\section{Conclusions}

Spectroscopic abundances for
6 elements in 325 stars show that, on average, solar neighbourhood
stars with large $|R_m - R_\odot|$ have lower abundances,
regardless of the chemical element considered. This relation is
interpreted as evidence of a well-defined age--abundance relation,
since stars with large $|R_m - R_\odot|$ would mainly be old. A
number of simulations of the orbits of disk stars born with
chemokinematical properties that are reasonable for the Galactic disk
confirm this interpretation.

The constraint provided by our data on the tightness of the AMR is 
relatively strong. Figure \ref{result} shows that a loose AMR, with 
$\sigma\approx 0.20$ dex, is still compatible with an  
arc-shaped $R_m$--$\langle{\rm [Fe/H]}\rangle$, albeit much less pronounced 
than what is observed. 

Our method highlights the possibility of studying the AMR and its dispersion through a 
technique that does not depend on age determination drawbacks. This could be achieved 
through a larger stellar sample and accurate (preferably, spectroscopic) abundances in order 
to better define the empirical $R_m$--$\langle{\rm [X/H]}\rangle$ relation.

\begin{acknowledgements}
We thank the suggestions by an anonymous referee that substantially improved this Letter. 
The authors acknowledge financial support by FAPERJ, FAPESP, CNPq,
PRONEX/FINEP, FUJB/UFRJ, and PIBIC/UFRJ. This research made use of the
SIMBAD database, operated by the CDS, Strasbourg, France.
\end{acknowledgements}


\end{document}